\documentstyle[12pt,epsf]{article}

\begin{document}

\title{
  \begin{flushright}
    \normalsize{hep-lat/9612019}\\
    \normalsize{UTHEP-354}\\
    \normalsize{December 1996}\\
  \end{flushright}
  \bigskip
  \bigskip
  \bigskip
  \bigskip
  \Large{\bf\mbox{Structure of Critical Lines in Quenched Lattice QCD}}\\
  \Large{\bf with the Wilson Quark Action}}
\author{Sinya Aoki, Tomoyuki Kaneda and Akira Ukawa\\
  \\
  \normalsize{\em Institute of Physics, University of Tsukuba}\\
  \normalsize{\em Tsukuba, Ibaraki 305, Japan}}
\date{}
\maketitle
\bigskip
\bigskip
\bigskip

\begin{abstract}
  The structure of critical lines of vanishing pion mass for the Wilson quark
  action is examined in quenched lattice QCD.  Numerical evidence is presented
  that the critical lines spread into five branches beyond $\beta=5.6$--$5.7$
  at zero temperature.  It is also shown that the critical lines disappear in
  the deconfined phase for the case of finite temperatures.
\end{abstract}

\newpage

\section{Introduction}

The problem of species doublers in lattice fermion formulations is well known.
The Wilson quark action, commonly used in numerical simulations of lattice QCD,
avoids this problem at the cost of explicit breaking of chiral symmetry.  This
means that the conventional Nambu-Goldstone mechanism associated with
spontaneous breakdown of chiral symmetry cannot be invoked to predict the
existence of massless pion for this quark action at a finite lattice spacing.
It has been found, however, that massless pion does appear with the Wilson
quark action once the bare quark mass $m_0$, or equivalently the hopping
parameter $K=1/(2m_0+8)$, is tuned appropriately.  It is now widely accepted
that the pion mass vanishes along a line $K=K_c(\beta)$, called the critical
line, which runs from $K_c(\beta=0)\approx1/4$ in the strong-coupling limit
$\beta=6/g^2=0$ to $K_c(\beta=\infty)=1/8$ in the weak-coupling limit
$\beta=\infty$ on the $(\beta,K)$ plane\cite{Kawamoto1981}.

One of us has suggested some years ago\cite{Aoki1984} that the massless pion
which appears for the Wilson quark action may be understood as a zero mode of a
second-order phase transition spontaneously breaking parity-flavor symmetry.
It has also been argued that, while there are only two critical lines in the
region of strong coupling, there exist ten critical lines in the weak-coupling
region, paired into five branches which reach the weak-coupling limit
$\beta=\infty$ at $M\equiv 1/2K=\pm4$, $\pm2$ and $0$.  Some
analytic\cite{Analytic} and numerical\cite{Numerical} results supporting this
view have been previously reported.

More recently arguments have been presented\cite{AokiUkawaUmemura1995} that the
end points of five branches move away from the weak-coupling limit for finite
temporal lattice sizes corresponding to finite temperatures, so that the
critical lines form five cusps at some $\beta\not=\infty$, above which no
critical line exists.  The existence of the second critical line paired with
the conventional one and a formation of a cusp by the two critical lines have
been confirmed numerically in a recent full QCD simulation at finite
temperature\cite{AokiUkawaUmemura1995}.

In this brief report we present further investigation of the structure of
critical lines for the Wilson quark action using quenched QCD.  One of the
motivations of the present work is to find evidence for the existence of five
pairs of critical lines toward weak coupling at zero temperature.  We expect
this structure to remain in the quenched approximation, in which case the
structure should be computationally much easier to confirm than in full QCD.
Another motivation is to examine how the structure of critical lines changes
for a finite temporal lattice size $N_t$ corresponding to finite temperature.
In particular we wish to examine how the first-order deconfinement phase
transition, which takes place at a fixed value $\beta=\beta_c(N_t)$ independent
of $K$ for quenched QCD, affects the critical lines.  Our expectation would be
that massless pion, and hence also the critical lines, disappears in the
high-temperature phase above the deconfinement transition $\beta>\beta_c(N_t)$.

\section{Results at zero temperature}

Finding the location of critical lines throughout the $(\beta,K)$ plane
requires a convenient indicator for their positions.  For this purpose we
employ the number of conjugate gradient iterations $N_{\rm CG}$ necessary to
invert the Wilson fermion matrix to a given value of the residual.  For
infinite volume this quantity increases toward critical lines and is expected
to diverge in between a pair of critical lines, where parity is spontaneously
broken, due to the presence of zero modes of the Wilson fermion matrix.  On a
finite lattice where zero mode eigenvalues should be slightly shifted away from
zero, we expect a peak in $N_{\rm CG}$ across each pair of critical lines.  Let
us note that the phase structure on the $(\beta,K)$ plane should be symmetric
under $K\to-K$.  For $K>0$ we thus expect three peaks in $N_{\rm CG}$ as a
function of $K$ toward weak coupling.

We have carried out a measurement of the number $N_{\rm CG}$ for a set of
values of $K>0$ at $\beta=5.0$, $5.5$, $5.6$, $5.7$ on a $6^4$ lattice, and at
$\beta=5.8$, $5.9$, $6.0$ on a $12^4$ lattice.  Results are obtained by an
average over $60$ (for $6^4$) or $20$ (for $12^4$) configurations separated by
$1000$ sweeps of the pseudo-heatbath algorithm.  The stopping condition is
taken to be $r=\sqrt{\|b-Dx\|^2/3V}<10^{-6}$ where $b$ represents a wall source
of unit strength for all space-time sites and colors, and $V=L^4$ the lattice
volume with $L=6$ or $12$.  Runs are made only for $K>0$ since the structure of
critical lines should be symmetric under $K\to-K$.

\begin{figure}
  \centerline{\epsfxsize=315.97pt \epsfbox{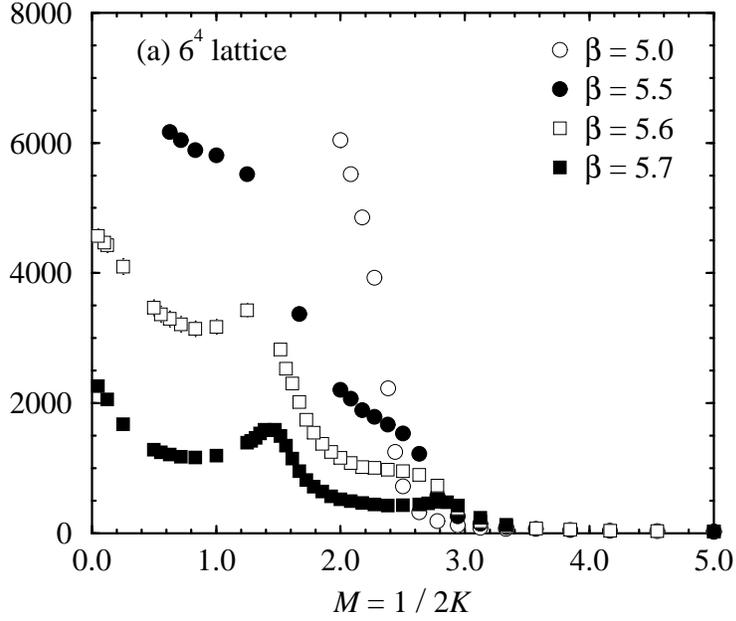}}
  \centerline{\epsfxsize=315.97pt \epsfbox{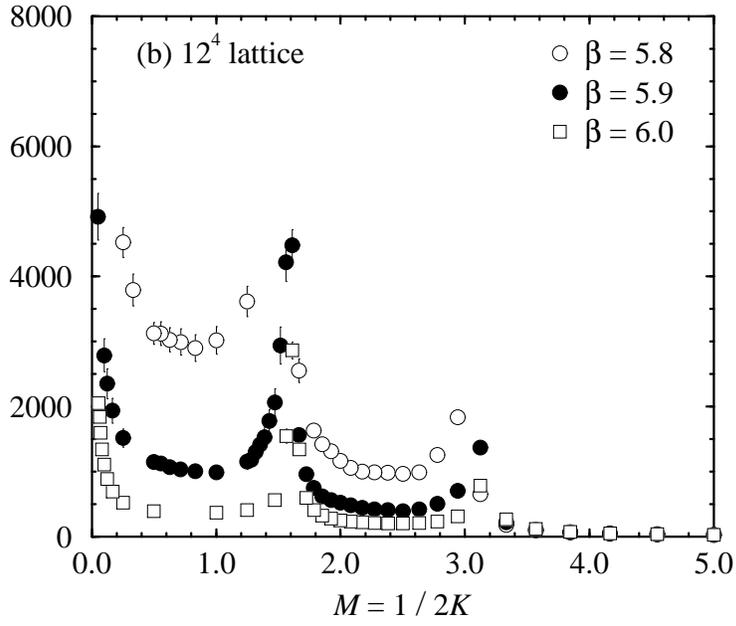}}
  \caption{Number of conjugate gradient iterations $N_{\rm CG}$ necessary to
    invert the Wilson fermion matrix with a wall source of unit strength for
    all sites and colors as a function of $M=1/2K$.  Stopping condition is
    chosen to be $r=\protect\sqrt{\|b-Dx\|^2/3V}<10^{-6}$ with $V=L^4$ the
    lattice volume.  (a)~$6^4$ lattice for $\beta=5.0$--$5.7$, (b)~$12^4$
    lattice for $\beta=5.8$--$6.0$.}
  \label{CG1}
\end{figure}

Our results are shown in fig.~\ref{CG1}(a) and (b).  On a $6^4$ lattice
(fig.~\ref{CG1}(a)) we observe appearance of three peaks at $M\approx3.0$,
$1.5$ and $0$ starting at $\beta=5.6$--$5.7$.  On a $12^4$ lattice
(fig.~\ref{CG1}(b)) the three peaks become increasingly sharper as $\beta$
increases.  For the peaks at $M\approx3.0$ and $1.5$, the peak position shifts
toward larger values of $M$, albeit only slightly over the range
$\beta=5.5$--$6.0$ examined in the simulation.  These results are consistent
with the expected structure of multiple critical lines appearing at
$\beta>5.6$--$5.7$.

In order to confirm that the peaks of $N_{\rm CG}$ observed at
$\beta>5.6$--$5.7$ correspond to five critical lines of massless pion, we have
calculated the pion mass at $\beta=6.0$ on a $10^3\times20$ and a
$16^3\times20$ lattice.  The pion propagator is averaged over $10$--$20$
configurations separated by $1000$ pseudo-heatbath sweeps.  The stopping
condition for quark propagator calculation is taken to be
$r=\|D^\dagger b-D^\dagger Dx\|/\|x\|<10^{-10}$ with $b$ a point source at the
origin.  We should note that we have encountered `exceptional' configurations
near the peaks of $N_{\rm CG}$, on which the pion propagator takes a W shape as
a function of time.  Since such exceptional configurations are never
encountered in our full QCD simulations\cite{AokiUkawaUmemura1995}, we think
that they are an artifact of the quenched approximation in which configurations
with small fermionic eigenvalues are not suppressed contrary to full QCD.  We
have therefore excluded the exceptional configurations from our propagator
average.

\begin{figure}
  \centerline{\epsfxsize=315.97pt \epsfbox{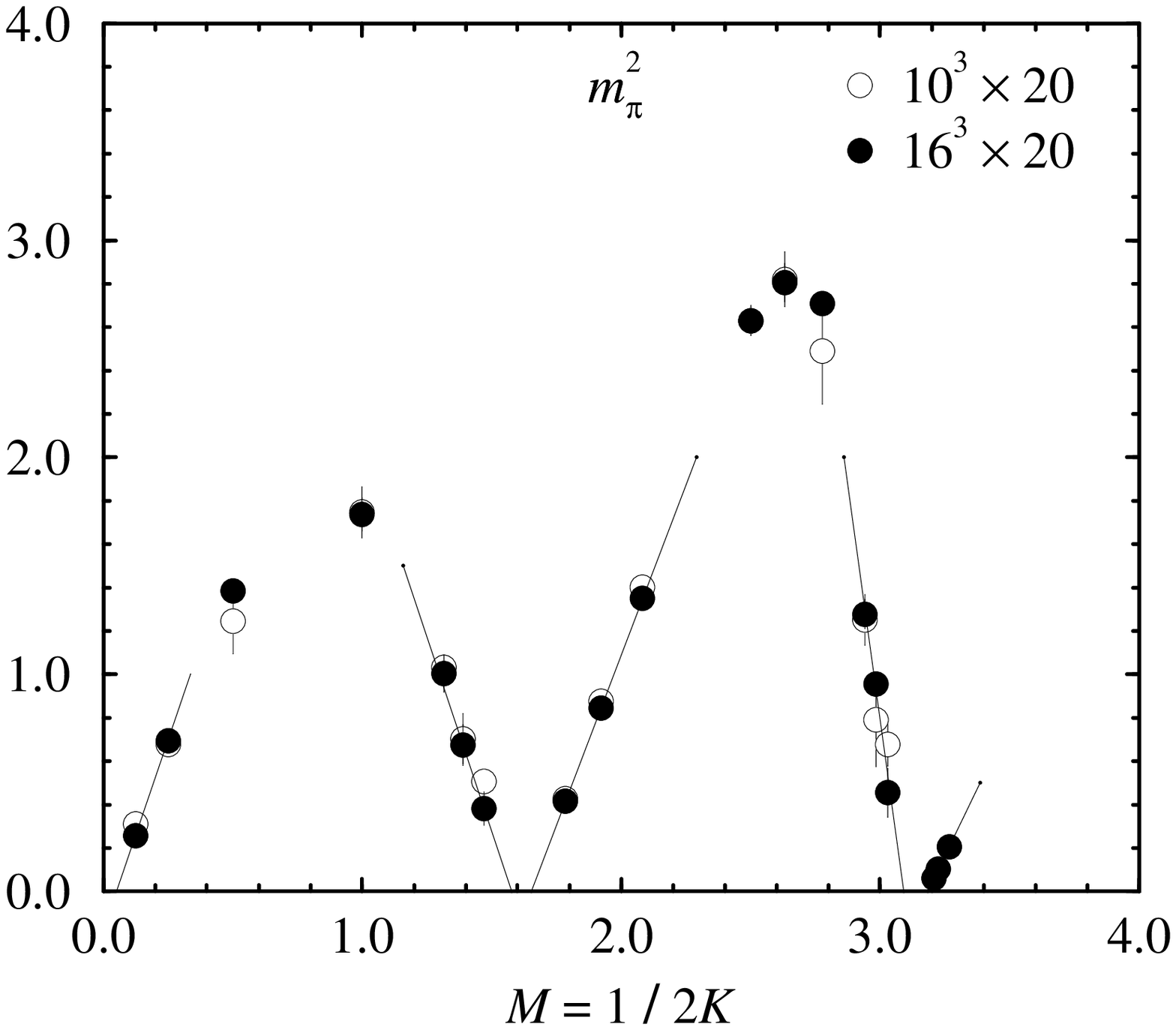}}
  \caption{Pion mass squared $m_\pi^2$ at zero temperature as a function of
    $M=1/2K$ at $\beta=6.0$.  Open circles are for a $10^3\times20$ lattice,
    and filled circles for a $16^3\times20$ lattice.  Lines show linear fits of
    $m_\pi^2$ in $M$.}
  \label{pmass6.0}
\end{figure}

We plot our results for the pion mass squared in fig.~\ref{pmass6.0} as a
function of $M=1/2K$ (see Table~1 for numerical values).  A
reasonable agreement of results for $10^3$ and $16^3$ spatial sizes shows that
finite spatial size effects are not severe in our data.  We clearly observe
that there exist four more critical values of vanishing pion mass in addition
to the conventional one located at the right most of the figure.  Making a
linear extrapolation of $m_\pi^2$ in $1/K$ using two or three data points, we
find $1/2K_c=0.052(14)$, $1.577(63)$, $1.653(5)$, $3.091(11)$ and $3.1848(44)$
for the five critical values from left to right.

\section{Results at finite temperature}

For a finite temporal lattice size $N_t$ corresponding to finite temperature,
the critical line should be defined by vanishing of the pion screening mass
extracted from the pion propagator for large spatial separations.  In quenched
QCD an interesting question is how the structure of critical lines changes
across the deconfinement transition which takes place at $\beta=\beta_c(N_t)$
independent of the hopping parameter $K$ of valence quark.  In the confining
phase $\beta<\beta_c(N_t)$, it is natural to expect that the structure of
critical lines remains qualitatively the same as at zero temperature except for
a possible shift of their location by a magnitude depending on $N_t$.  On the
other hand, the pion screening mass would not vanish in the deconfined phase
$\beta>\beta_c(N_t)$, and hence critical lines would disappear.  For values of
$K$ close to the conventional critical value, such a behavior has been seen in
a previous work\cite{iwasakituboi90}.  Our aim here is to carry out a
systematic study over the entire interval of the hopping parameter.

Our simulation is made for two temporal lattice sizes $N_t=4$ and $8$.  The
critical coupling is known to be
$\beta_c(N_t=4)=5.69226(41)$\cite{FukugitaOkawaUkawa1990} and
$\beta_c(N_t=8)=6.0609(9)$\cite{Bielefeld96}.  For $N_t=4$, runs are made at
$\beta=5.635$ and $5.75$ on a $12^3\times4$ lattice, and for $N_t=8$ at
$\beta=5.90$ and $6.10$ on a $16^3\times8$ lattice.  Quark propagators are
computed on a periodically doubled lattice $12^2\times24\times4$ in the former
case and on the same lattice $16^3\times8$ for the latter.  Numerical results
for the pion screening mass are listed in Table~1.

\begin{figure}
  \centerline{\epsfxsize=315.97pt \epsfbox{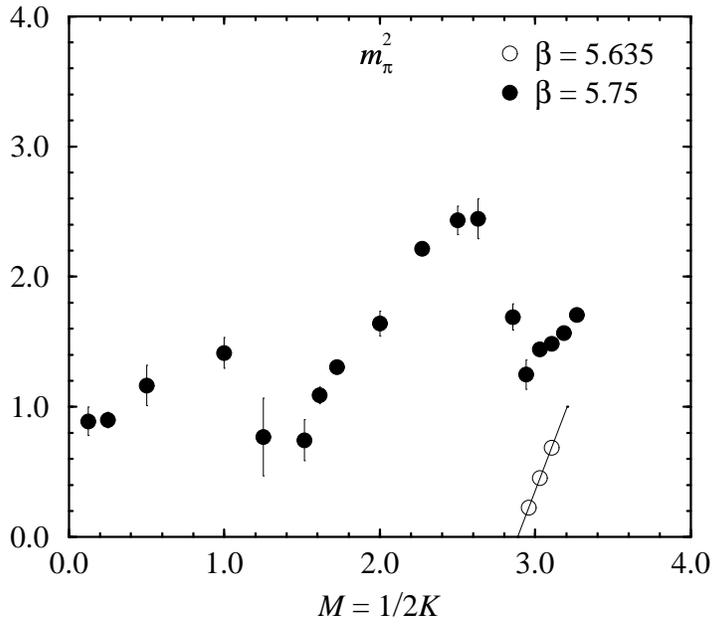}}
  \caption{Pion screening mass squared $m_\pi^2$ as a function of $M=1/2K$ on
    a $12^2\times24\times4$ lattice.  Open circles are for $\beta=5.635$, and
    filled circles for $\beta=5.750$.  Line shows a linear fit of $m_\pi^2$ in
    $M$.}
  \label{pmassNt4}
\end{figure}

In fig.~\ref{pmassNt4} we plot by open circles the pion mass squared on a
$12^2\times24\times4$ lattice as a function of $M=1/2K$ at $\beta=5.635$, which
is in the confining phase.  We find only a single critical point located at
$1/2K_c=2.886(17)$.  Exploring the region $K>K_c$ at $K=0.2$--$0.4$, we found
that the gauge configurations are dominated by exceptional ones for which the
conjugate gradient solver for quark propagator takes over $5000$--$10000$
iterations to converge, while for $K<K_c$ a thousand iterations or less are
sufficient.  We take this as an indication that the parity-broken phase extends
over $\infty>K>K_c$ at $\beta=5.635$ for $N_t=4$, {\em i.e.}, the system is
still in the strong coupling region where only one critical line exists for
$K>0$.

For $\beta=5.75$, which is in the deconfined phase, results for the pion
screening mass squared are shown by filled circles in fig.~\ref{pmassNt4}.
While an overall pattern of $m_\pi^2$ as a function of $M=1/2K$ is similar to
the zero-temperature case shown in fig.~\ref{pmass6.0}, the pion mass does not
vanish for any value of $K$ for the present case.  Thus the critical lines are
absent in the deconfined phase as expected.

\begin{figure}
  \centerline{\epsfxsize=315.97pt \epsfbox{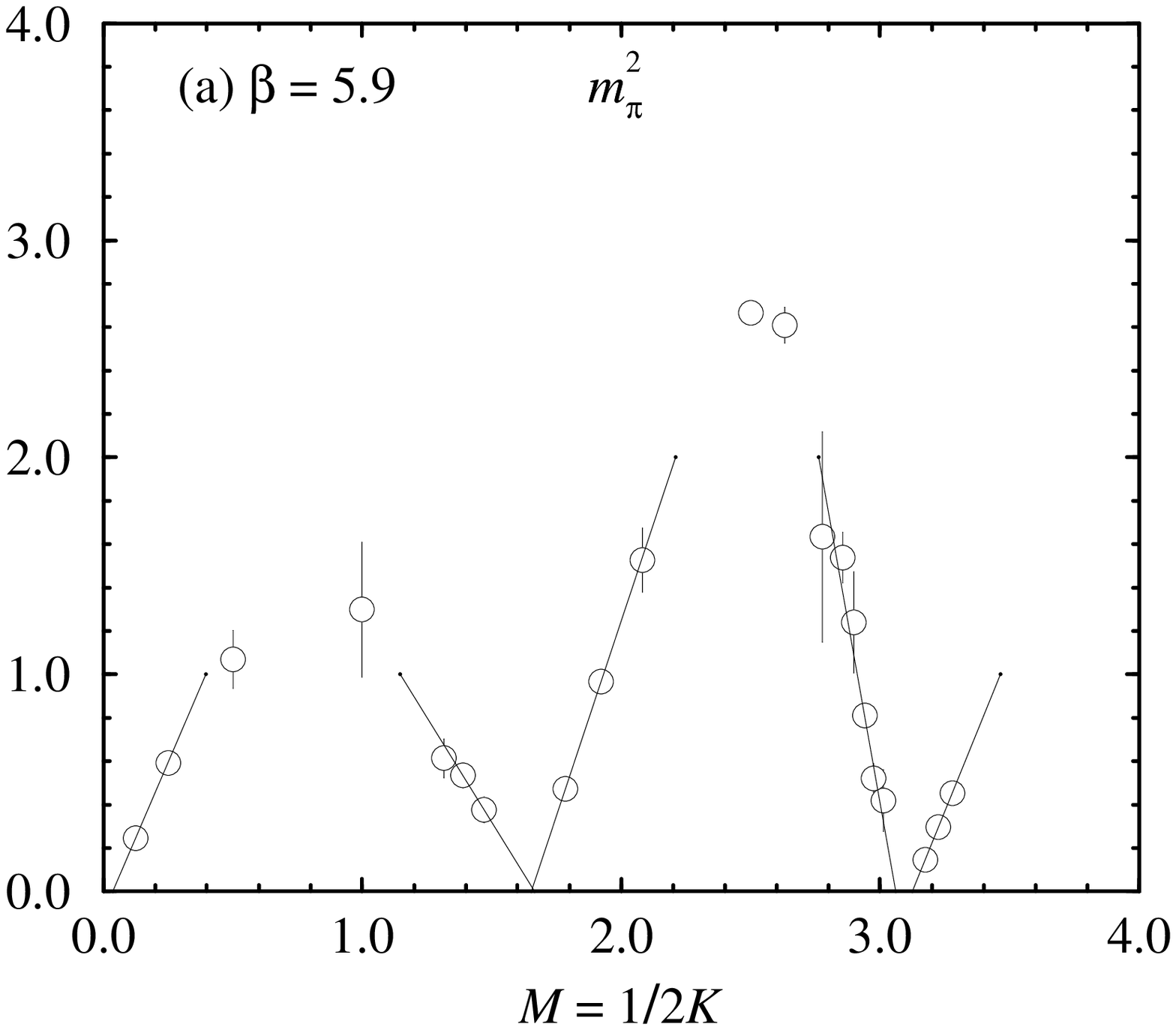}}
  \centerline{\epsfxsize=315.97pt \epsfbox{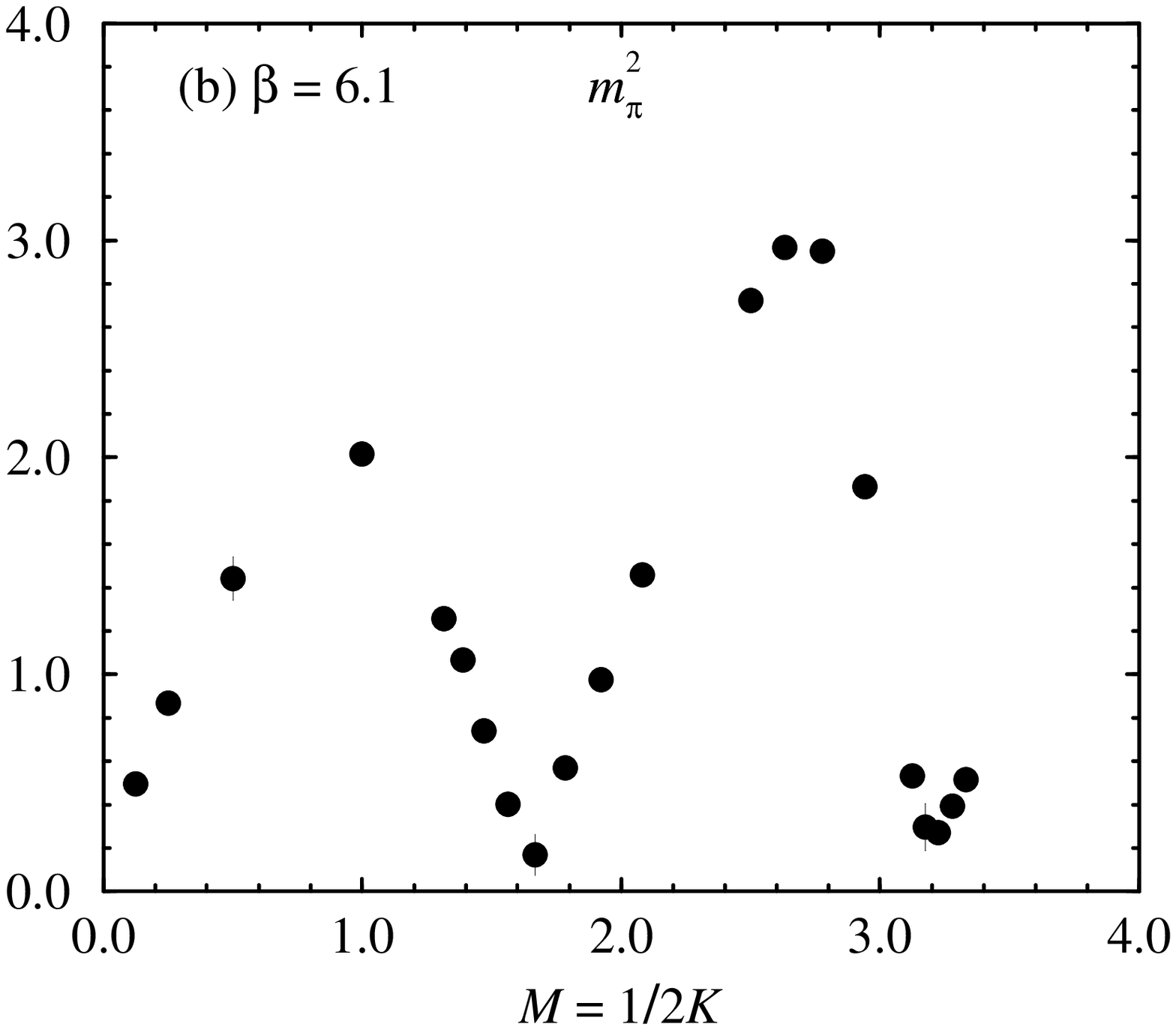}}
  \caption{Pion screening mass squared $m_\pi^2$ as a function of $M=1/2K$ on a
    $16^3\times8$ lattice.  (a)~$\beta=5.9$, (b)~$\beta=6.1$.  Lines in (a)
    show linear fits of $m_\pi^2$ in $M$.}
  \label{pmassNt8}
\end{figure}

Our results for $N_t=8$ at $\beta=5.9$ (confined phase) is plotted in
fig.~\ref{pmassNt8}(a).  We observe two critical values at $1/2K\approx3.0$.
The pion mass also appears to vanish at $1/2K\approx1.65-1.66$ and
$\approx0.04$, although our data is not sufficiently precise to resolve if
there are two critical values separated by a narrow gap or $m_\pi^2$ has a
small but non-vanishing value.  In fig.~\ref{pmassNt8}(b) we show how the
behavior changes at $\beta=6.1$ (deconfined phase).  We clearly observe
non-vanishing of pion mass in the deconfined phase.

\section{Conclusion}

In this brief report we have presented numerical evidence for the existence of
five pairs of critical lines beyond $\beta=5.6$--$5.7$ in quenched QCD at zero
temperature, including an estimate of five values of $K_c$ at $\beta=6.0$.

It has recently been suggested that the two critical lines forming each pair
merge at a finite value of $\beta$, turning into a single line\cite{creutz}.
Our results show that this possibility is not realized up to $\beta=6.0$ in
quenched QCD.

We have also shown that the critical lines are absent above the deconfinement
phase transition $\beta>\beta_c(N_t)$.  Since the deconfinement transition is
of first order, we expect the critical lines to be sharply cutoff at the
critical value $\beta=\beta_c(N_t)$.  Our data, however, is not precise enough
to examine this point in detail.

\section*{Acknowledgements}

Numerical simulations for this work have been carried out on VPP500/30 at
Science Information Processing Center, University of Tsukuba, and at Center of
Computational Physics, University of Tsukuba.  This work is supported in part
by the Grants-in-Aid of the Ministry of Education (Nos. 04NP0801, 08640349,
08640350).

\newpage

\begin{figure}
  \centerline{\epsfxsize=446.85pt \epsfbox{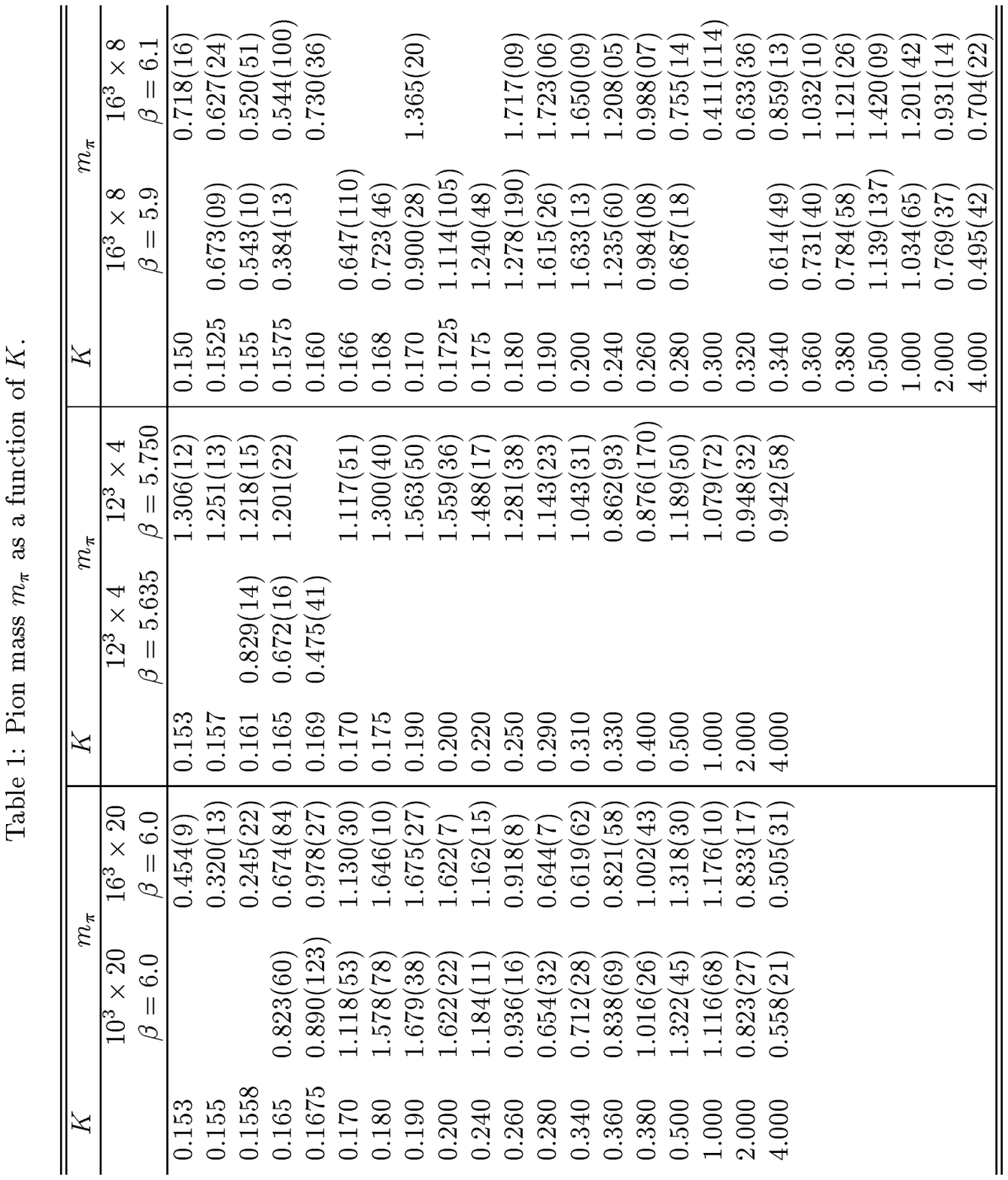}}
  \label{tab:pmass}
\end{figure}

\end{document}